\documentclass[aps,prl,twocolumn,superscriptaddress]{revtex4-1}
\usepackage{amsmath}
\usepackage{epsf}
\usepackage{braket}
\usepackage{amsfonts}
\usepackage{amssymb}
\usepackage{graphicx}
\usepackage{hyperref}
\usepackage[all]{hypcap}
\usepackage{subfigure}
\hypersetup{backref,
	pdfpagemode=FullScreen,
	colorlinks=true, linkcolor=blue, urlcolor=blue, citecolor=blue}
\usepackage[sort&compress]{natbib}
\begin{document}
\title{Mixing of coherent waves on a single three-level artificial atom} 

\author{T. H\"{o}nigl-Decrinis}
\email[]{Teresa.Hoenigl-Decrinis.2014@rhul.ac.uk}
\affiliation{Physics Department, Royal Holloway, University of London, Egham, Surrey TW20 0EX, United Kingdom}
\affiliation{National Physical Laboratory, Teddington, TW11 0LW, United Kingdom}

\author{I.V. Antonov}
\affiliation{Physics Department, Royal Holloway, University of London, Egham, Surrey TW20 0EX, United Kingdom}
\affiliation{National Physical Laboratory, Teddington, TW11 0LW, United Kingdom}

\author{R. Shaikhaidarov}
\affiliation{Physics Department, Royal Holloway, University of London, Egham, Surrey TW20 0EX, United Kingdom}
\affiliation{Moscow Institute of Physics and Technology, 141700 Dolgoprudny, Russia}

\author{V.N. Antonov}
\affiliation{Physics Department, Royal Holloway, University of London, Egham, Surrey TW20 0EX, United Kingdom}
\affiliation{Moscow Institute of Physics and Technology, 141700 Dolgoprudny, Russia}
\affiliation{Skolkovo Institute of Science and Technology, Nobel str. 3, Moscow, 143026, Russia}

\author{A.Yu. Dmitriev}
\affiliation{Moscow Institute of Physics and Technology, 141700 Dolgoprudny, Russia}

\author{O.V. Astafiev}
\email[]{Oleg.Astafiev@rhul.ac.uk}
\affiliation{Physics Department, Royal Holloway, University of London, Egham, Surrey TW20 0EX, United Kingdom}
\affiliation{National Physical Laboratory, Teddington, TW11 0LW, United Kingdom}
\affiliation{Moscow Institute of Physics and Technology, 141700 Dolgoprudny, Russia}
\date{\today}
%\pacs{42.50.Gy, 42.65.Ky}

\begin{abstract}
We report coherent frequency conversion in the gigahertz range via three-wave mixing on a single artificial atom in open space. All frequencies involved are in vicinity of transition frequencies of the three-level atom. A cyclic configuration of levels is therefore essential, which we have realised with an artificial atom based on the flux qubit geometry. The atom is continuously driven at two transition frequencies and we directly measure the coherent emission at the sum or difference frequency. Our approach enables coherent conversion of the incoming fields into the coherent emission at a designed frequency in prospective devices of quantum electronics.
\end{abstract}

\maketitle
For a long time research in experimental quantum optics focused on studying ensembles of natural atoms~\citep{Miller:2005uu, Walther:2006da}. However, there have been huge advances in performing analogous quantum optics experiments using other systems~\citep{Hanson:2007wx, Dutt:2007ug, deRiedmatten:2008ck}. In particular, superconducting artificial atoms are remarkably attractive to study quantum optics phenomena. The artificial atoms are nano-scale electronic circuits that can be fabricated using well established techniques and can therefore be easily scaled up to larger systems. Their energy levels can be engineered as desired, and strong coupling can be achieved with resonators and transmission lines~\citep{Hoi:2013dr, Hoi:2011tn, Astafiev:2010cm, VanLoo:2013wm}.
This greater control of parameters allows one to reproduce quantum optics phenomena with improved clarity or even reach regimes, that are unattainable with natural atoms. For instance coherent population trapping~\citep{Kelly:2010hj}, electromagnetically induced transparency~\citep{Murali:2004ib, AbdumalikovJr:2010vv}, Autlers-Townes splitting~\citep{Autler:1955gb, Baur:2009ee, Sillanpaa:2009db, Novikov:2013jt, Cho:2014wy}, and quantum wave mixing~\citep{Dmitriev:2017vd} have been experimentally observed in superconducting three-level systems~\citep{Inomata:2014eh, Dumur:2015di, Li:2015gz, Bianchetti:2010ih, Sathyamoorthy:2014jz}. Moreover, three-level atoms can be used to cool quantum systems~\citep{You:2008jd, Valenzuela:2006hp}, amplify microwave signals~\citep{Astafiev:2010cc} and generate single or entangled pairs of photons~\citep{PhysRevLett.119.140504} -- important applications for future quantum networks. Here we investigate three-wave mixing, a nonlinear optical effect that can occur in cyclic three-level atoms, which are lacking in nature~\citep{Liu:2014db}, but can easily be realised with superconducting artificial atoms. The only suitable natural systems for the three-wave mixing are chiral molecular three-level systems without inversion symmetry~\citep{Patterson:2013jn}. However, these systems cannot be tuned in frequency. 
Different to Josephson junction based parametric three-wave mixing devices~\citep{Roch:2012gy}, that rely on mixing on a classical non-linearity, we implement here another method to generate three-wave mixing using a single cyclic or $\Delta$-type artificial atom. This was considered theoretically in references~\citep{Liu:2014db, Zhao:2017ho}.

We directly measure the coherent emission of the cyclic three-level atom under two external drives corresponding to two atomic transitions. The emission occurs at a single mixed frequency (sum or difference). This emission is a corollary of coherent frequency conversion but inherently differs from classical frequency conversion~\citep{Boyd:2003wk, Scully:1999uy} which would result in sidebands at the sum and difference frequencies.
Previously, coherent atomic excitations using two frequencies have been studied in a single dc-SQUID phase qubit circuit with two internal degrees of freedom~\citep{Lecocq:2012gv}. However, in this work, we realise coherent frequency conversion with a cyclic artificial atom in open space, which offers some advantages over placing it in a cavity. In particular, it allows to directly detect the coherent (elastic) component of the emitted field at sum or difference frequencies of the artificial atom~\citep{AAAbdumalikov:2011ko}. %by a vector network analyser (VNA)
This work establishes innovative quantum electronics that enables three-wave mixing, and coherent frequency conversion.

\begin{figure}[!htb]
\centering
\includegraphics[width=1\columnwidth]{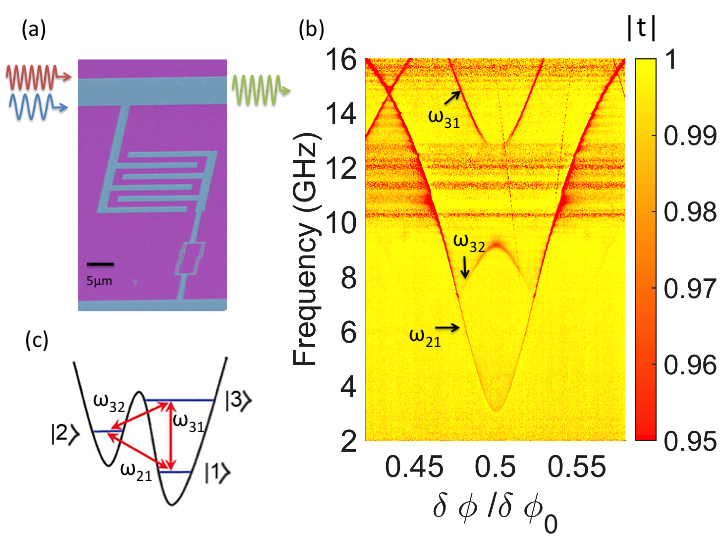}
\caption{(a) False-coloured micrograph of the device taken at an angle. The three-level artificial atom consisting of a superconducting loop with four Josephson junctions is capacitatively coupled to the transmission line. (b) Spectroscopy of the single artificial three-level atom. To detect all three transition frequencies we overdrive the atom and measure transmission as a function of flux bias and a probe driving frequency. (c) A cyclic-type artificial atom away from the degeneracy point $\delta\Phi\neq 0$  with transition frequencies $\omega_{21}/2\pi = 6.48$ GHz, $\omega_{32}/2\pi = 8.35$ GHz, and $\omega_{31}/2\pi = 14.83$ GHz.}
\label{fig:1}
\end{figure}

Our device consists of a superconducting loop ($\sim 10\mu m^{2} $) interrupted by four Josephson junctions. This geometry is based on the flux qubit~\citep{Mooij1036} where one of the Josephson junctions, the $\alpha$-junction, has a reduced geometrical overlap by a factor of $\alpha$. It is capacitatively coupled to a 1D transmission line via an interdigitated capacitance of $C=6$ fF (see Fig~\ref{fig:1}(a)) resulting into a photon rate in the range from several MHz to a few tens of MHz depending on frequency. The device parameters (Josephson energy $E_{J}/h=65$ GHz, charging energy ($E_{C}=e^2/2C$) $E_{C}/h=19$ GHz, and $\alpha=0.45$) have been chosen such that the three lowest transition frequencies fall into the frequency measurement band of our experimental setup. The coupling to the transmission line is strong enough so that non-radiative atom relaxation is negligible and hence the majority of photons from the atom are emitted into the transmission line.
The device was fabricated by means of electron-beam lithography and shadow evaporation technique with controllable oxidation.

The transition frequencies, $\omega_{12}$, $\omega_{23}$, and $\omega_{13}$ are controlled by the external magnetic flux threaded through the loop, $\Phi=\Phi_{0}/2+\delta\Phi$, where $\Phi_{0}$ is the flux quantum and $\delta\Phi$ is the detuning from the energy degeneracy point of the artificial atom.
The atomic transition energies are found by performing transmission spectroscopy using a vector network analyser (VNA). We sweep the frequency of a probe microwave against the flux bias $\delta\Phi$, as seen in Fig.~\ref{fig:1}(b). For the  $|3\rangle \rightarrow |2\rangle$ transition to be clearly visible in spectroscopy we overdrive the artificial atom. The working point is set away from the degeneracy point $\delta\Phi\neq 0$, where all transitions are allowed, forming a cyclic or $\Delta$-type atom with transition frequencies $\omega_{21}/2\pi = 6.48$ GHz, $\omega_{32}/2\pi = 8.35$ GHz, and $\omega_{31}/2\pi = 14.83$ GHz ($\omega_{31} = \omega_{21}+\omega_{32}$), as schematically shown in Fig.~\ref{fig:1}(c).

\begin{figure}[!tb]
\centering
\includegraphics[width=1\columnwidth]{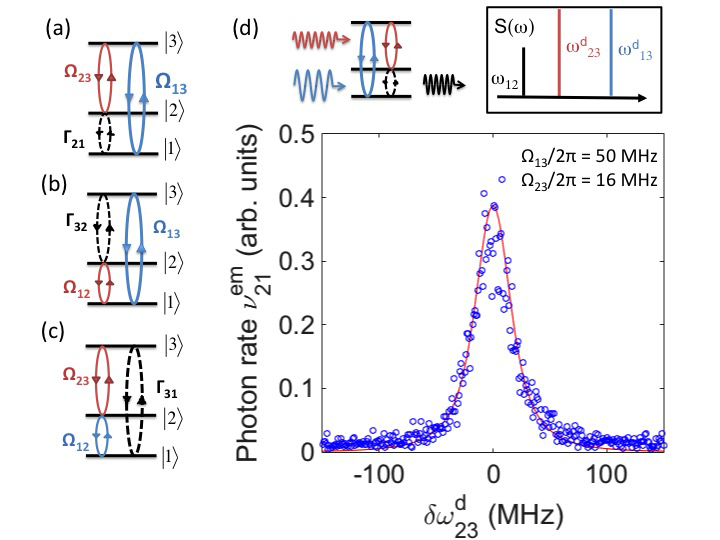}
\caption{Coherent emission in a driven three-level atom with energy diagrams of the pumping schemes. (a) The three-level atom is continuously driven with driving amplitudes $\Omega_{23}$ and $\Omega_{13}$, in b) with driving amplitudes $\Omega_{12}$ and $\Omega_{13}$, and in c) with driving amplitudes $\Omega_{12}$ and $\Omega_{23}$. d) The measured coherent emission peak at $\omega_{12}$ in terms of photon rate, $\nu_{21}^{em}$, under driving amplitudes $\Omega_{23}/2\pi=16$ MHz, $\Omega_{13}/2\pi=50$ MHz, as a function of detuning of the driving frequency, $\delta\omega_{23}^{d}$. The inset schematically shows the typically measured spectrum.}
\label{fig:2}
\end{figure}

The experiment is performed in a dilution refrigerator at base temperature $T=12$ mK, at which point thermal excitations are suppressed and negligible. 
We investigate coherent emission of the three-level artificial atom under two continuous drives. All regimes shown in Fig.~\ref{fig:2}(a-c) have been measured with different driving field amplitudes, $\Omega_{ij}$, between states $|i\rangle$ and $|j\rangle$, where $i$ and $j$ are 1, 2 or 3.
Quantum mechanics dictates that there can only be emission at a frequency corresponding to an atomic transition within the atom. 

To understand the physical process, it is instructive to use the second quantisation approach. For the interaction of waves on the single quantum system only one scattering process can occur at the same instant. Introducing creation (annihilation) operator, $a^{\dagger}_{ij}$ $(a_{ij})$ of a photon at frequency $\omega_{ij}$, the allowed multi-photon processes, limited by the transitions of the atom, are described by $a_{31}a_{32}^{\dagger}a_{21}^{\dagger}$, and $a_{31}^{\dagger}a_{32}a_{21}$.
These two processes conserve energy and explain the creation of the field in Fig.~\ref{fig:2}(a-c) denoted as dashed black lines.

First, let us focus on the case when transitions $\ket{1}\rightarrow\ket{3}$ and $\ket{2}\rightarrow\ket{3}$ are driven with excitation frequencies $\omega_{31}^{d}=\omega_{31}+\delta\omega_{31}$, $\omega_{32}^{d}=\omega_{32}+\delta\omega_{32}$  (Fig.~\ref{fig:2}(a)) and emission at $\ket{2}\rightarrow\ket{1}$ is measured.
Here $\delta\omega_{ij}$ are small detunings from their corresponding atomic transition frequencies $\omega_{ij}=\omega_{i}-\omega_{j}$ with $i>j$. 
In the rotating wave approximation of the semi-classical picture, the three-level artificial atom under two drives $\omega_{31}^{d}$, $\omega_{32}^{d}$ coupling the atomic states through the dipole interaction $\hbar\Omega_{ij}=\phi_{ij}V_{ij}$, with $\phi_{ij}$ the atomic dipole moment, is described by the Hamiltonian

\begin{equation}
\begin{aligned}
H={}&-\hbar(\delta\omega_{31}\sigma_{11}+\delta\omega_{23}\sigma_{22})\\
&-\hbar\left[\frac{\Omega_{13}}{2}(\sigma_{13}+\sigma_{31})+\frac{\Omega_{23}}{2}(\sigma_{32}+\sigma_{23})\right],
\end{aligned}
\end{equation}
where $\sigma_{ij}=\ket{i}\bra{j}$ is the transition operator.
The dynamics of the system are governed by the Markovian master equation.

The atom interacting with 1D open space emits a coherent wave~\citep{Astafiev:2010cm, AAAbdumalikov:2011ko}
\begin{equation}
V^{em}_{ji}(x,t)=i\frac{\hbar\Gamma_{ji}}{\phi_{ji}}\braket{\sigma_{ij}}e^{i (k_{ji}|x|-\omega_{ji}t)}
\label{eq:V}
\end{equation}
where $\braket{\sigma_{ij}}=\rho_{ji}$ is found from the stationary solution ($\dot\rho=0$) of the master equation.
The spectral density $S(\omega)=\frac{1}{2\pi}\int_{-\infty}^{+\infty}\langle\hat{V}^{em}_{ij}(0)\hat{V}_{ji}^{em}(\tau)\rangle_{ss}e^{i\omega \tau} d\tau$, where the subscript ($ss$) of the correlator denotes the stationary solution, decomposes into incoherent and coherent parts ~\citep{Carmichael:1991}. Using a spectrum analyser we monitor the narrow emission peak, corresponding to the coherent component of the emission $S_{coh}=\hbar \omega Z_{0} \Gamma_{ji} \langle \sigma_{ij} \rangle_{ss} \langle \sigma_{ji}\rangle_{ss} \delta(\omega-\omega_{ij})$ with the impedance of the transmission line $Z_0$ and where we have substituted $\Gamma_{ji} = \frac{\omega Z_{0}}{\hbar \phi_{ji}}$ ~\citep{Astafiev:2010cm}. The narrow peak power (mathematically a delta function) in the emission spectrum is expected to be

\begin{equation}
P(\omega)=\frac{\hbar\omega\Gamma_{ji}}{2} |\braket{\sigma_{ij}}|^2,
\end{equation}
where $P=\frac{|V_{ji}^{em}|^2}{2Z_0}$. Here $\omega$ is in the vicinity of  the transition frequency $\omega_{21}/ 2\pi=6.48$ GHz as schematically shown above Fig.~\ref{fig:2}(d). The narrow peak power $P(\omega)$ can be measured by a spectrum analyser or any equivalent methods (see Supplementary Methods). Further we refer to the Voltage amplitude $V^{em}$, which can be extracted from $P(\omega)$, as the coherent emission. The linewidth of the emission peak is as narrow as the linewidths of the generator emission that is driving the artificial atom, indicating that the phase does not fluctuate more than the phase of the generator and therefore indicating coherency. We expect the phase, which has not been measured here, to be locked to the difference or sum frequency of the pumps as confirmed in the simulations. %We explain simulations later

The general analytic solution of the generated field is bulky. In the approximation of weak driving regime, the solution simplifies to $\langle\sigma_{ij}\rangle\approx\frac{\Omega_{ik}}{\lambda_{ik}}\frac{\Omega_{kj}}{\lambda_{kj}}$ where $\lambda_{mn}=\gamma_{mn}-i\delta\omega_{mn}$ (all indices take values 1,2, or 3). However, to efficiently generate the mixed wave the atom has to be strongly driven ($\Omega_{mn}\gg\gamma_{mn}$). 
If our device is used as a single side band mixer, the maximum power it would sustain is limited by the relaxation time of the transition and must be $\leq\hbar\omega\Gamma_{ij}/8$ since $|\langle\sigma_{ji}\rangle| \leq 1/2$. Due to the operating principle, the bandwidth of such a device is restrained by the transition frequencies of the cyclic atom.
\begin{figure}[!htb]
\centering
\includegraphics[width=1\columnwidth]{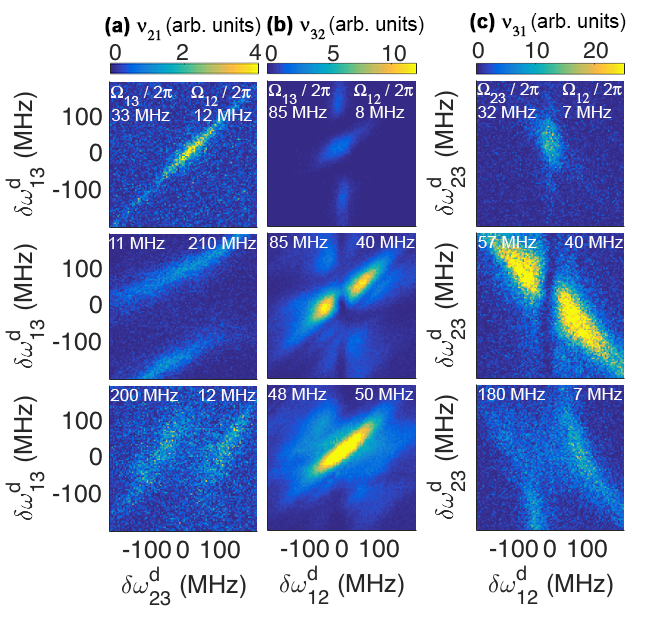} 
\caption{Measurement of coherent emission expressed as photon rate $\nu^{em}$ in arbitrary units as a function of frequency detuning $\delta\omega^{d}$ of the two drives with amplitudes $\Omega_{ij}$ indicated on the panels. (a) Emitted photon rate of the transition from $\ket{2}\rightarrow \ket{1}$, $\nu^{em}_{21}$, with Rabi frequencies corresponding to the respective field strengths $\Omega_{13}$, $\Omega_{23}$. (b) Emitted photon rate of the transition from $\ket{3}\rightarrow \ket{2}$, $\nu^{em}_{32}$, with Rabi frequencies corresponding to the respective field strengths $\Omega_{13}$, $\Omega_{12}$. (c): Emitted photon rate of the transition from $\ket{3}\rightarrow \ket{1}$, $\nu^{em}_{31}$, with Rabi frequencies corresponding to the respective field strengths $\Omega_{12}$, $\Omega_{23}$.}
\label{fig:3}
\end{figure}

Fig.~\ref{fig:2}(d) shows the measured coherent emission peak as a function of detuning of the driving frequency, $\delta\omega_{23}^d$, expressed as photon rate (in arb. units), 
$\nu^{em}_{21}=\frac{P(\omega)}{\hbar\omega}$ under weak pumping amplitudes ($\Omega_{13}<<\gamma_{13}$, $\Omega_{23}<<\gamma_{23}$, where $\gamma_{ij}$ are dephasing rates). Note that we are measuring only the elastically scattered coherent emission from the atom. Each point in Fig.~\ref{fig:2}(d) corresponds to the narrow emission peak exemplified as series of dotted peaks.

\begin{figure}[!hbt]
\centering
\includegraphics[width=1\columnwidth]{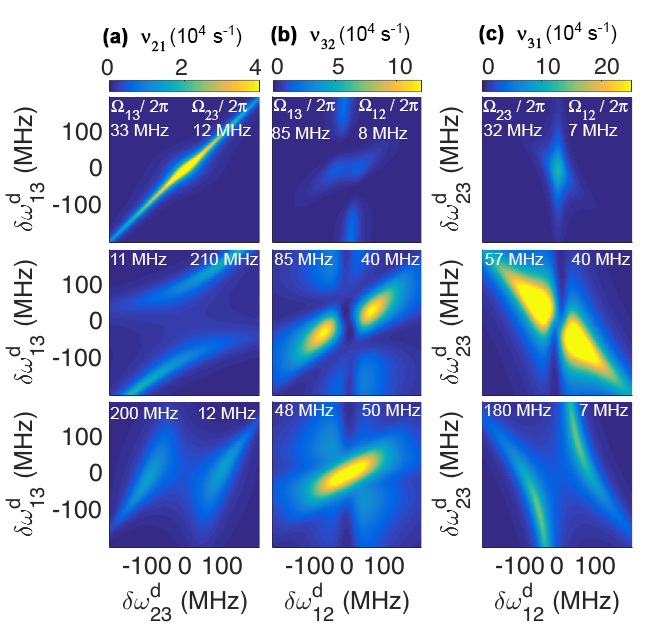} 
\caption{Numerical simulations of coherent emission expressed as photon rate $\nu_{em}$ as a function of frequency detuning $\delta\omega^{d}$ of the two drives with amplitudes $\Omega_{ij}$ indicated on the panels. (a) Emitted photon rate of the transition from $\ket{2}\rightarrow \ket{1}$, $\nu^{em}_{21}$, with Rabi frequencies corresponding to the respective field strengths $\Omega_{13}$, $\Omega_{23}$. (b) Emitted photon rate of the transition from $\ket{3}\rightarrow \ket{2}$, $\nu^{em}_{32}$, with Rabi frequencies corresponding to the respective field strengths $\Omega_{13}$, $\Omega_{12}$. (c) Emitted photon rate of the transition from $\ket{3}\rightarrow \ket{1}$, $\nu^{em}_{31}$, with Rabi frequencies corresponding to the respective field strengths $\Omega_{12}$, $\Omega_{23}$.}
\label{fig:4}
\end{figure}

We then measure the coherent emission as a function of detuning of $\delta\omega_{23}^d$ for varying values of $\Omega_{13}$, while keeping $\Omega_{23}$ constant. Splitting of the coherent emission under large driving amplitude $\Omega_{13}$ is observed which appears due to level splitting induced by driving fields.
This splitting is investigated further by recording the coherent emission versus detuning of the two drives for various combinations of powers. As seen in Fig.~\ref{fig:3}(a), the direction of the splitting is determined by the stronger drive: $\Omega_{13}>>\Omega_{23}$ leads to $\Omega_{13}$ splitting level $\ket{1}$; $\Omega_{23}>>\Omega_{13}$ leads to $\Omega_{23}$ splitting level $\ket{2}$ and the splitting pattern in the coherent emission is turned by 90 degrees. %refer to coherent emission as$V_{21}^{em}$ or photon rate as in Fig.?

In an analogous way, we pump transitions between states $\ket{3}$ and $\ket{1}$ with driving frequency $\omega_{31}^{d}=\omega_{31}+\delta\omega_{31}$ and transitions between states $\ket{2}$ and $\ket{1}$ with driving frequency $\omega_{21}^{d}=\omega_{21}+\delta\omega_{21}$ (Fig.~\ref{fig:2}(b)) resulting in the Hamiltonian
\begin{equation}
\begin{aligned}
H={}&-\hbar(\delta\omega_{21}\sigma_{22}+\delta\omega_{31}\sigma_{33})\\
&-\hbar\left[\frac{\Omega_{12}}{2}(\sigma_{12}+\sigma_{21})+\frac{\Omega_{13}}{2}(\sigma_{32}+\sigma_{23})\right].
\end{aligned}
\end{equation}

In this pumping scheme, the emission power of the coherent emission of transitions between states $\ket{3}$ and $\ket{2}$, $V^{em}_{32}$, is extracted and a narrow peak in the power spectrum at $\omega_{32}/2\pi=8.35$ GHz is recorded.

The coherent emission between states $\ket{3}$ and $\ket{2}$ is monitored as a function of detuning of the drives, $\delta\omega_{13}^d$ and $\delta\omega_{12}^d$, for several combinations of driving amplitudes, $\Omega_{13}$ and $\Omega_{12}$, Fig.~\ref{fig:3}(b), the result being more complex than in the previous driving configuration. It becomes apparent that the coherent emission from the atom depends on all relaxation and dephasing rates. The bright coherent emission line stretching diagonally from the bottom left to the top right corner in Fig.~\ref{fig:3}(b) is primarily determined by dephasing on the $\ket{3}$ to $\ket{2}$ transition, $\gamma_{23}$. The vertical coherent emission line that appears for some combinations of powers strongly depends on the dephasing rate $\gamma_{12}$. Emission lines broaden when the two driving frequencies are comparable to each other and larger than their dephasing rates. The understanding of the dependency of parameters on the emission line was developed from experimental data and numerical simulations.  In contrast to the experiment, all input parameters of the numerical simulations can be varied independently.

To achieve coherent frequency upconversion we pump transitions between states $\ket{2}$ and $\ket{1}$ with driving frequency $\omega_{21}^d = \omega_{21}+\delta\omega_{21}$ and transitions between states $\ket{3}$ and $\ket{2}$ with driving frequency $\omega_{32}^d=\omega_{32}+\delta\omega_{32}$, see Fig.~\ref{fig:2}(c). The Hamiltonian for this configuration is
\begin{equation}
\begin{aligned}
H={}&-\hbar(\delta\omega_{21}\sigma_{11}+\delta\omega_{32}\sigma_{33})\\
&-\hbar\left[\frac{\Omega_{12}}{2}(\sigma_{12}+\sigma_{21})+\frac{\Omega_{23}}{2}(\sigma_{32}+\sigma_{23})\right].
\end{aligned}
\end{equation}
As expected, we observe a single narrow coherent emission peak in the emission power spectrum only at the sum frequency $\omega_{12}/2\pi+\omega_{23}/2\pi=14.83$ GHz, but not at the difference frequency $\omega_{12}-\omega_{23}$, confirming that our results cannot be explained by mixing with a classical nonlinearity.
Similar to the previous pumping configurations, the coherent emission peak is split under a strong driving amplitude. Fig.~\ref{fig:3} (c) shows the behaviour of the coherent emission $V^{em}_{13}$ expressed as photon rate $\nu_{31}$ as a function of detuning of the drives $\delta\omega_{12}/2\pi$ and $\delta\omega_{23}/2\pi$ for a range of driving powers.

Finally, we numerically simulate our experimental results using the master-equation formalism with the Lindblad term
\begin{equation}
\begin{aligned}
L[\rho]={}&(\Gamma_{31}\rho_{33}+\Gamma_{21}\rho_{22})\sigma_{11}+(\Gamma_{32}\rho_{33}-\Gamma_{21}\rho_{22})\sigma_{22}\\
&-(\Gamma_{31}\rho_{33}+\Gamma_{23}\rho_{22})\sigma_{33}-\sum_{i\neq j} \gamma_{ij}\rho_{ij}\sigma_{ij}.
\end{aligned}
\end{equation}
Here $\gamma_{ij}=\gamma_{ji}$ is the damping rate of the off-diagonal terms (dephasing) and $\Gamma_{ij}$ is the relaxation rate between the levels $\ket{i}$ and $\ket{j}$.
In the numerical simulations dephasing and relaxation rates are arbitrary numbers. The constraints are that dephasing and relaxation rates are fixed by our sample (the three-level atom) throughout the experiment, and the input driving powers are varied but known (we set them at the generators). By finding the correspondence between the simulations, Fig.~\ref{fig:4}, and our measurement results, Fig.~\ref{fig:3}, we extract $\Gamma_{21}/2\pi=8$ MHz, $\gamma_{21}/2\pi=8$ MHz, $\Gamma_{32}/2\pi=38$ MHz, $\gamma_{32}/2\pi=42$ MHz, $\Gamma_{31}/2\pi=41$ MHz, and $\gamma_{31}/2\pi=39.5$ MHz agreeing with our expectations. %expectations from what? Spectroscopy, device design
%In the numerical simulations we chose $\Gamma_{21}/2\pi=8$ MHz, $\gamma_{21}/2\pi=8$ MHz, $\Gamma_{32}/2\pi=38$ MHz, $\gamma_{32}/2\pi=42$ MHz, $\Gamma_{31}/2\pi=41$ MHz, and $\gamma_{31}/2\pi=39.5$ MHz which give the best correspondence between the experiment, Fig.~\ref{fig:3}, and simulations, Fig.~\ref{fig:4}.

%Those numbers are arbitrary in our simulations. They are chosen such that the simulations reproduce all our experimental results that were measured for different set of driving powers. The constraints are that dephasing and relaxation rates have to be equal across all measurements, and the input driving powers are known (we set them at the generators). By finding the correspondence between the simulations and our measurement results, we extract the dephasing rates at relaxation rates very accurately and they do correspond to our expectations.

Since our measurement set-up has not been pre-calibrated, experimental results include gain and attenuation coefficients and are therefore presented in arbitrary units. Comparing the experimental results with the numerical simulations would yield a calibration of our output line. This calibration depends on frequency and was not the focus of this work. Nevertheless, we obtain a calibration factor of our output line of $G_{21}=2*10^5$, $G_{32}=1.3*10^6$, $G_{31}=10^5$ for each of the three transition frequencies, $6.48$ GHz, $8.35$ GHz, $14.83$ GHz respectively. The visible difference between our experimental measurement results, Fig.~\ref{fig:3}, and the numerical simulations, Fig.~\ref{fig:4}, are noise in the experiment.

In conclusion, we have demonstrated three-wave mixing and coherent frequency conversion using a single cyclic three-level artificial atom. The fundamental difference from classical Josephson junction based parametric three-wave mixing devices~\citep{Roch:2012gy} is that here transition frequencies of the artificial atom are mixed to generate a single coherent emission peak at the sum or difference frequency. A requirement for this phenomena to occur is a cyclic-type atom, which is absent in nature due to electric-dipole selection rules, but can easily be realised with superconducting artificial atoms. Thus we suggest a unique method of generating coherent fields at a designed frequency by mixing on the single artificial atom.
 
%{\it{Acknowledgements.}}
We acknowledge the Russian Science Foundation (grant N 16-12-00070) for supporting the work.
\bibliographystyle{apsrev4-1}
\bibliography{Ref}
\end{document}